# Extending the W3C Data Privacy Vocabulary to Cover the German Health Data Use Act - Lessons Learned

Julian FLAKE[a,1], Christian HENKE[b], Alessandra S. KUNTZ[b], Matthias LÖBE[c], Harshvardhan J. PANDIT[d] and Christian DRAEGER[c]
[a] *Institute for Software Technology (IST), University of Koblenz, Germany*
[b] *Department of Medical Informatics, University Medical Center Göttingen (UMG), Göttingen, Germany*
[c] *Institute for Medical Informatics, Statistics and Epidemiology (IMISE), University of Leipzig, Germany*
[d] *Trinity College Dublin, Ireland*
ORCiD ID: JF 0009-0006-4623-1463, CH 0000-0002-4541-4018, ASK 0000-0002-8259-2577, ML 0000-0002-2344-0426, HJP 0000-0002-5068-3714 and CD 0000-0001-8835-4548

**Abstract.** Finding, sharing, and reusing data requires common interoperable vocabularies to adequately describe machine-actionable data policies. Semantic Web standards are a proven way to express such information describing data access and reuse. Narrative descriptions get a stable and agreed meaning by formal definitions of terms that have a globally unique identifier. The Data Privacy Vocabulary (DPV) represents a community standard for privacy, data protection, and related topics. DPV uses modular namespaces to define jurisdictions, such as "EU", and to represent concepts from laws within that jurisdiction, such as the General Data Protection Regulation (GDPR) and the European Health Data Space (EHDS). In this work, we present the creation of a DPV extension representing the German Health Data Use Act (Gesundheitsdatennutzungsgesetz (GDNG)) that has been set into force in March 2024. The GDNG regulates sharing and using health data for secondary use, such as research purposes and policymaking. It builds the German legal foundation for the future EHDS as the European initiative and regulation providing rules for sharing of health data within Europe. The parallel development of the EHDS and GDNG extensions resulted in many shared concepts, which contributed to an overarching extension: the DPV sector health extension. It sets the foundation of aligned privacy terms for future EHDS member states looking to build a national vocabulary aligned with the EHDS. All extensions were released as draft extensions with the latest DPV version 2.3, released 25th of February 2026.

[1] Corresponding Author, Julian Flake, Institute for Software Technology (IST), University of Koblenz, Universitätsstr. 1, 56070 Koblenz, Germany; E-mail: flake@uni-koblenz.de.

# 1. Introduction

## 1.1. Background

Application of the German Health Data Use Act (Gesundheitsdatennutzungsgesetz (GDNG)) in research contexts faces conceptual and technical challenges, particularly regarding consistent machine-readable representation of relevant information. A central issue is the lack of stable, technically referenceable identifiers for terms within the German legal framework.

Semantic Web standards, such as the Resource Description Framework (RDF) and the Open Digital Rights Language (ODRL), provide foundations for machine-readable data policies. However, they currently lack coverage of national regulations like the GDNG and specific constructs such as opt-out models. Furthermore, data policies should rely on consistent references to legal bases, e.g., through policy URLs, but are hindered by heterogeneous, non-standardized referencing schemes. Even the Federal Law Gazette (Bundesgesetzblatt) only partially implements the European Legislative Identifier (ELI) and does not support paragraph-level referencing. The technical representation of German state-level legislation varies widely, in part no stable identifiers are provided.

In contrast, the Data Privacy Vocabulary (DPV), which is a community standard developed by the W3C Data Privacy Vocabularies and Controls Community Group (DPVCG), is designed to enable interoperable machine-readable descriptions of legal terms, supporting integration into existing legal and technical ecosystems. This approach has been successfully applied to the General Data Protection Regulation (GDPR) and other EU regulations, including the EU's Data Governance Act and AI Act [1,2,3].

## 1.2. Objective and Requirements

The DPV formally defines terms, such as legal bases, from regulatory frameworks such as the GDPR by first defining a 'core' vocabulary and then representing the GDPR concepts as an extension. To represent GDNG as a vocabulary, a similar extension would be necessary. This follows linked data principles that recommend reuse of existing concepts, and to constrain them to represent additional context, such as the specific legal bases or entities defined in the GDNG. This means rather than forking DPV and bending established concepts to the German use case, we create a compatible extension that integrates national concepts from GDNG into the existing DPV landscape. By doing this, we take advantage of the DPVCG's open governance structure that allows interested parties to propose additional concepts and to develop new use cases. Fortunately, as the DPVCG was already working on an extension for the EHDS, working on a national legal legislation with a largely similar scope was seen as mutually beneficial. A key requirement for the GDNG extension is therefore its alignment with the EHDS concepts.

# 2. State of the art

## 2.1. Related Work

The lack of a formalized vocabulary for defining concepts and relationships in the area of data protection in general and GDPR in particular motivated the development of DPV in 2018. As the vocabulary matured, it became clear that GDPR concepts also exist similarly in other legislations' scopes and as higher level concepts independent of specific legislation or jurisdiction. Therefore, several concepts were generalized to create a core namespace called 'DPV' from which extensions were created to represent laws such as GDPR and subjective taxonomies for personal data categories and locations [4].

For DPV 2.0, the scope was revised, and additional extensions were created to represent specific topics and taxonomies, such as risk management, Artificial Intelligence (AI), additional laws such as the EU AI Act. Guides on the usage of DPV were also included in the 2.0 release [5].

### *2.2. Shortcomings*

DPV extensions were not built for national level legislation alone. Other approaches towards machine readability of legal terms and context often involve creating custom code sets, such as HL7 FHIR's Consent Policies and ValueSets[2]. While these approaches support local or use-case specific implementations, they do not provide globally governed, reusable identifiers for legal concepts across laws and jurisdictions. This is necessary because health-data uses are highly likely to also require addressing concepts from the GDPR, such as consent, as well as sectorial concepts, such as specific medical purposes. Therefore, the use of DPV helps address this shortcoming through a reference set of terms and legal basis with corresponding IRIs, that still can be used within FHIR structures, e.g. Consent.policyBasis, while being aligned across regulations by the DPVCG.

## 3. Concept

To implement a GDNG extension in the DPV, consensus on the scope of work must first be reached within the DPVCG and among the persons involved in the extension development. The process begins with a formal request submitted to the DPVCG, such as through the mailing list. This request is discussed in one of the weekly meetings, where it is evaluated whether the proposed topic falls within the scope of the DPV. If accepted, the DPVCG assigns a suitable namespace, in this case `/dpv/legal/de/gdng`, and a corresponding suggested prefix (`de-gdng`).

We then initiated discussions to understand the law, its scope, and relevant sources, primarily relying on the official publication by the German Federal Government. In the case of the GDNG, the corresponding Federal Law Gazette with the title that translates to "Law for improved usage of health data" consists of six articles [6]. While only Article 1 of the Federal Law Gazette covers the GDNG itself, articles 2 to 5 amend four already existing German laws. Article 6 defines that the whole package is set into force on the day of announcement. We set the scope of this publication to Article 1 only, i.e., the nine sections of the GDNG itself. The nine sections are enumerated § 1 to § 9.

[2] https://hl7.org/fhir/consent.html

To support understanding, we created models[3] of existing and fictitious processes related to the GDNG. These models help to clarify relevant participants and their interactions, as well as data artifacts and flows. They illustrate which concepts in the extension describe which process artifacts, facilitating discussions about the domain and helping identification of concepts to be represented.

As a tool to collect and discuss identified concepts and corresponding data, we used a Google spreadsheet. It includes columns for the term itself (the machine-readable identifier), a status (e.g., "proposed", "accepted"), a label (intended to be human-readable), a definition, a source (e.g., to refer to a specific paragraph of the regulation), the contributing persons, a date representing when it was finalized, and additional usage hints. It also included the columns *ParentType* and *ParentTerm* to represent relations such as subclasses and hierarchies between concepts within the same extension and across DPV extensions.

## 4. Implementation

### *4.1. Solution / Results*

During the first iteration, the spreadsheet was filled with identified terms, defined labels, and definitions derived from the GDNG in German, along with references to their exact sources and dates of their additions. Later, English variants for the values of these three columns were defined in three additional columns to meet the default language requirements of the DPVCG. The remaining columns of the table were filled in an ad hoc fashion and updated iteratively.

Following DPVCG guidelines, we defined modules that group related concepts. For the GDNG draft extension, we identified the *Basic Concepts*, *Actors, Query and Access*, *Data Linkage,* and *Lawfulness* as the modules. For each module, a separate sheet was added to the spreadsheet document. Capturing related concepts allowed us to connect the new concepts to existing ones, primarily hierarchically. We defined relations between concepts of the GDNG extension to existing concepts in the DPV core vocabulary, as well as concepts in other DPV extensions, for example, the `eu-gdpr` extension or the `sector-health` extension [7]. As an example, Figure 1 shows how three different concepts of `SecondaryUse` co-exist in three extensions and how they relate to each other. A cube represents a DPV extension, a rectangle represents a specific concept, and an arrow from concept `A` to concept `B` represents the existence of two relationships: `A rdfs:subClassOf B` and `A skos:broader B`.

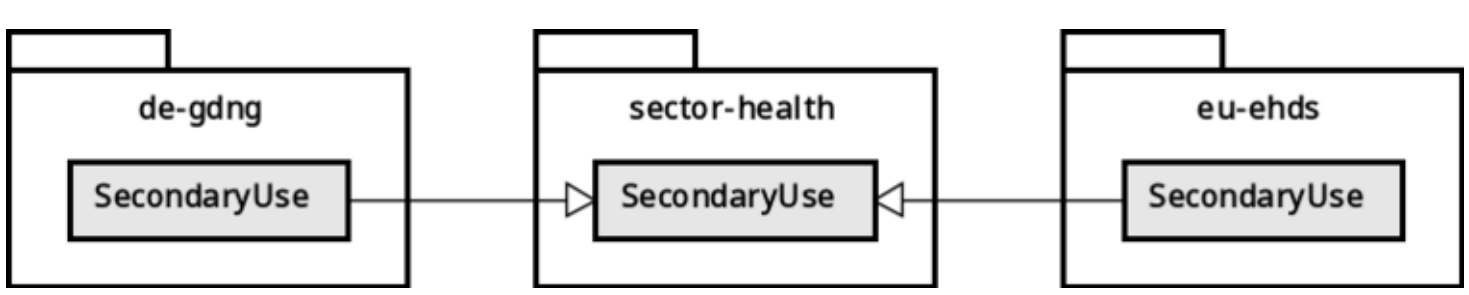

**Figure 1.** Relations between concepts of different DPV extensions

As one can see, there now exist three `SecondaryUse` concepts, which is intentional. The `sector-health` extension defines concepts common to the health domain without any specific reference to a particular regulation. The term `sector-health:SecondaryUse` defines this concept generally, loosely described as "secondary use of health data". This distinguishes this concept from secondary use of data in other contexts than the health domain, for example, for fraud detection [8]. The term `de-gdng:SecondaryUse` further narrows the concept to "secondary use of health data in the sense of the GDNG". The definition of `sector-health:SecondaryUse` is additionally useful to link the GDNG and other vocabularies like DPV's EHDS extension, which also defines a narrower secondary use concept of the `sector-health:SecondaryUse` concept, `eu-ehds:SecondaryUse`, which can be described as "secondary use of health data in the sense of EHDS regulation". By these separate concepts and their relations, legal interpretation of the two regulations is avoided, and we don't claim that secondary use of health data in one regulation has the same meaning as secondary use of health data in the other regulation. But also, if such a relation were (re-)defined in the future, then it could be added without changing the concepts themselves, e.g., to state that GDNG's secondary use is the same as EHDS's secondary use. During the development of the GDNG extension, several such concepts were identified that now co-exist in both GDNG and EHDS extensions. For these concepts, we created additional generalized concepts to represent a common parent within the sector-health extension.

[3] https://w3c-cg.github.io/dpv/2.3/legal/de/gdng/#vocab-de-gdng

After several iterations of discussing, defining, describing, reformulating, and translating concepts, a draft HTML document representing the GDNG extension was created. With the exception of the translation columns, we mainly followed the process and formats already established by the DPVCG, which mainly consists of three steps:

I. Download the Google spreadsheets, create one CSV file per sheet
II. Create machine-readable RDF representations from CSV data
III. Create human-readable HTML representations from RDF.

Since these steps can happen in an automated pipeline (e.g., a GitHub Action), an iterative approach was applied for extracting the CSVs from the online spreadsheet (Step I), converting them to RDF formats, e.g., Turtle, JSON-LD (Step II), and generating the HTML documentation containing each concept's documentation by using Jinja2[4] as a template engine (Step III). The Jinja2 template created for the GDNG extension directly contains or refers to additional artifacts like tables, diagrams, and explanatory text

While the machine-readable serializations contain labels and definitions for German and English (defined as `skos:prefLabel "Sekundärdatennutzung"@de, "Secondary Use"@en`), we decided to create two human-readable documents, one for each language, containing the same amount of information about the 54 identified terms (see [9]), to enable users to access the documentation in their preferred language.

### *4.2. System in Use*

The Open Digital Rights Language (ODRL) is a W3C standard ('Recommendation' in W3C terminology) for expressing policies and defining fine-grained data use permissions [10]. While ODRL is a generic specification with a limited vocabulary, it is not designed for the needs for health data or even scientific data in general. Existing work has demonstrated the use of DPV with ODRL to address this issue [11].

[4] https://jinja.palletsprojects.com/en/stable/

Figure 2 gives a short example policy that defines rules for an X-ray image data transfer. It defines that `ex:UniversityClinicLeipzig` assigns to `ex:InstituteForSoftwareTechnologyKoblenz`, which is in that context a `de-gdng:ThirdParty` according to the GDNG, the permission to use a received X-ray image for the purpose of `de-gdng:MedicalRehabNursingResearch` in the sense of GDNG § 6(1) point 2 and provided the use is covered by consent in the sense of GDNG § 6(3) (`de-gdng:P6-3-Consent`[5]). It is possible to add further conditions, e.g., if the patient excluded some purposes.

```
ex:UniversityClinicLeipzig a de-gdng:DataProcessingHealthcareFacility .
ex:InstituteForSoftwareTechnologyKoblenz a de-gdng:HealthDataUser .
ex:x-rayImageAB a de-gdng:HealthData .

ex:policy a odrl:Set ;
    odrl:permission [
        odrl:target ex:x-rayImageAB ;
        odrl:assigner ex:UniversityClinicLeipzig ;
        odrl:assignee ex:InstituteForSoftwareTechnologyKoblenz ;
        odrl:action odrl:use ;
        odrl:constraint [
            odrl:leftOperand odrl:Purpose ;
            odrl:operator odrl:isA ;
            odrl:rightOperand de-gdng:MedicalRehabNursingResearch ] ;
        odrl:constraint [
            odrl:leftOperand dpv:LegalBasis ;
            odrl:operator odrl:isA ;
            odrl:rightOperand de-gdng:P6-3-Consent ] ] .
```

**Figure 2.** Example ODRL Policy in Turtle syntax with terms of the GDNG DPV extension. For better clarity, the namespace definitions have been omitted.

This example demonstrates that we are still not able to express all rules that apply to this data processing for secondary use purposes. The GDNG not only defines permissions to process data, it also defines obligations. It requires the registration of research projects in a primary registry in the WHO registry network and the publication of research results. This could perfectly be formulated as an ODRL duty within the same policy, as the example permission is formalized. However, there is currently no concept that captures the exact registration or publication obligations laid down by the GDNG. The lack of such concepts is a key reason for declaring the DPV extension as still being in draft stage. The topic of GDNG §8, the duty to register research projects using health data and to publish their results, will be covered by upcoming releases.

## 5. Lessons learned

A major challenge for the DPVCG and for DPV adopters, which includes an international audience, was the German-only nature of GDNG legal text. This was the first case of a DPV extension developed for a regulation that could not use an English text as primary source. We (all native German speakers) read the German regulation, extracted terms, labels, and definitions in German first. We decided to provide translations for all concepts (terms, labels, definitions, and usage hints) and used resources to assist us such as German governmental websites that provided an English translation for certain named entities. In most cases, translations were manually created. In the end, the German terms (the machine-readable IRIs) were not used.

[5] https://w3id.org/dpv/2.3/legal/de/gdng/#P6-3-Consent

During the vocabulary development process, we tried several approaches for machine assisted translation. For example, the European Commission provides AI language tools announced as free and secure[6]. Initially, we tried to use an automatically translated version of the GDNG. An initial inspection revealed that the results were not satisfying in many cases. Furthermore, we did not have any evidence to assess whether a translation is an official translation, an established translation, or an arbitrary translation that may be valid in some contexts, but not valid in legal contexts. We decided to rely on manual translations instead, followed by a discussion of the resulting terms with English native speakers and technical and legal experts of the DPVCG. The resulting vocabulary was iteratively refined in this manner with formal validation, including inter-rater agreement and coverage assessment, as identified future work.

One difficulty we faced was following the DPVCG convention of formulating the IRIs such that it clearly indicates a reference to a specific paragraph in the regulation. As the GDNG uses notations typical for German legal text, the § sign represents a 'paragraph' for GDNG whereas EU regulations would typically use 'article'. This mismatch led to confusion while discussing the regulation with the DPVCG in English. An example of a DPV IRI referring to a specific GDPR legal basis is `eu-gdpr:A6-1-b`, which, according to the traditionally applied schema in DPV, easily translates to "GDPR Art. 6(1)b", which in turn translates to "GDPR Article 6, paragraph 1, point b". In Germany, the outermost unit of text inside a regulation is typically called "Paragraph" rather than "Artikel". Therefore, the IRIs in GDNG extension that refer to a certain paragraph start with a "`P`" and not with an "`A`" as in the example `de-gdng:P6-3-Consent`. The "`P`" is followed by the reference to the named paragraph (German "Absatz"), which is enumerated with "(3)". In this particular case, the third sub unit of the sixth paragraph (§ 6 (3)) describes three different legal bases. Which is why the IRI needs the additional suffix "`-Consent`" to refer to one of the three specific legal bases of "§ 6(3)". The resolution of crafting IRIs for GDNG in this manner also helped the DPVCG refine its methodology for representing legal concepts beyond typical EU-level regulations.

## 6. Conclusion

The GDNG represents an important development for health data use in Germany. We provide the first instance of representing GDNG concepts in an interoperable and machine-readable format by extending the DPV as a community standard. Our work, which is published as a draft specification in DPV version 2.3, sits alongside other drafts representing the EHDS and a common sector-health extension that will ease integration of other national legislations into the DPV. To demonstrate its use, we have provided an ODRL example that shows the use of our work to represent data use policies.

[6] https://translation.ec.europa.eu/tools-and-resources/ai-translation-and-language-tools

Future work will focus on further refining and validating the draft vocabulary, including based on community feedback and supporting its future application in interoperable policy descriptions, including its application in interoperability scenarios such as ODRL and HL7 FHIR. Full coverage of all GDNG obligations is required to bring the draft into a stable release. Since all three extensions, GDNG extension, EHDS extension and sector-health extension are released as drafts, feedback on them is very welcome. Future work also needs to address potential changes to the GDNG legislation, as ministries within Germany are currently working on the national legal framework for the EHDS, which likely includes changes or additions to existing laws.

We hope that this case report serves as an encouragement to others seeking to establish a standardized vocabulary for national legislation. Our work shows how a joint effort can be mutually beneficial, both to a national standardization effort as well as the multinational community group. Especially the depth of legal translations proved challenging and the consolidated effort of the DPVCG was of great help. Overall, the quality of the GDNG vocabulary provided could not have been achieved without the extensive support and insightful discussion in the DPVCG.

## Declarations

*Conflict of Interest:* The authors declare that there is no conflict of interest.

*Contributions of the authors:*
Writing: All; Concept: CD, JF; Extraction of GDNG Terms: JF, CD, CH; Alignment with EHDS Terms: JF, CD, ML, HP; Implementation: JF, HP; Application in ODRL: CH, JF; Proofreading & Revision: All

*Acknowledgement:* Research was supported by DFG grants number 315072261 and number 442326535. The contributions of Julian Flake received funding from EU's Digital Europe Programme, grant number 101123471 (EDGE-Skills), and from EU's Horizon Europe Framework Programme, grant number 101129822 (TITAN EOSC). The contributions of Harshvardhan J. Pandit are funded by the RECITALS Project under EU's Horizon program with grant agreement No. 101168490.

## References

[1] Esteves B, Rodríguez-Doncel V. Analysis of ontologies and policy languages to represent information flows in GDPR. Semantic Web. 2024 May 14;15(3):709–43. doi:10.3233/sw-223009.

[2] Pandit, H. J., Lindquist, J., & Krog, G. P. (2024). Implementing ISO/IEC TS 27560: 2023 consent records and receipts for GDPR and DGA. In Annual Privacy Forum (pp. 228-251). Cham: Springer Nature Switzerland. doi:10.1007/978-3-031-68024-3_12.

[3] Golpayegani, D., Pandit, H. J., O'Sullivan, D., & Lewis, D. (2026). Semantic frameworks to support implementation of the EU AI Act. Computer Law & Security Review, 61, 106315. doi:10.1016/j.clsr.2026.106315.

[4] Pandit, H.J. *et al.* (2019). Creating a Vocabulary for Data Privacy. In: Panetto, H., Debruyne, C., Hepp, M., Lewis, D., Ardagna, C., Meersman, R. (eds) On the Move to Meaningful Internet Systems: OTM 2019 Conferences. OTM 2019. Lecture Notes in Computer Science, vol 11877. Springer, Cham. doi:10.1007/978-3-030-33246-4_44.

[5] Pandit J., Esteves B., Krog, G.P., Ryan, P., Golpayegani, D., Flake, J. (2025). Data Privacy Vocabulary (DPV) – Version 2.0. In: Demartini, G., *et al.* The Semantic Web – ISWC 2024. ISWC 2024. Lecture Notes in Computer Science, vol 15233. Springer, Cham. doi:10.1007/978-3-031-77847-6_10.
[6] Federal Law Gazette Part I structure number 102, dd. 26 March 2024, https://www.recht.bund.de/eli/bund/bgbl-1/2024/102, last accessed 2026-03-31.
[7] Esteves B, Pandit H.J., Flake J (2026) Health Sector Concepts (SECTOR-HEALTH), Draft Community Group Report 25 February 2026, https://w3c-cg.github.io/dpv/2.3/sector/health/, last accessed 2026-03-31.
[8] Solove, DJ., A Taxonomy of Privacy. University of Pennsylvania Law Review, Vol. 154, No. 3, p. 477, January 2006, GWU Law School Public Law Research Paper No. 129, doi:10.2307/40041279
[9] Draeger C, Henke C, Flake J, Löbe M (2026), Legal Concepts for German Gesundheitsdatennutzungsgesetz (DE), Draft Community Group Report 25 February 2026, https://w3c-cg.github.io/dpv/2.3/legal/de/gdng/, last accessed 2026-03-31.
[10] Renato I., Serena V., ODRL Information Model 2.2, W3C Recommendation 15 February 2018, https://www.w3.org/TR/odrl-model/, last accessed 2026-03-31
[11] Esteves, B., Pandit, H. J., Rodríguez-Doncel, V. (2021), ODRL profile for expressing consent through granular access control policies in solid. In 2021 IEEE European Symposium on Security and Privacy Workshops (EuroS&PW) (pp. 298-306). IEEE. doi:10.1109/EuroSPW54576.2021.00038.